\documentstyle[12pt]{article}
\begin{document}
\begin{titlepage}
{\hfill$\vcenter{ \hbox{TP-USl/96/09}
                \hbox{December 1996}}$ }
\vspace{2cm}

\begin{center}
{\Large\bf Heavy neutrinos production and decay \\
in future $e^+e^-$ colliders} \\
\vspace{1.5 cm}
{\large\bf J. Gluza}\footnote{e-mail address:
gluza@us.edu.pl} and
{\large\bf M. Zra{\l}ek}\footnote{e-mail address:
zralek@us.edu.pl} \\
\vspace{ 0.5 cm}
Department of Field Theory and Particle Physics \\
Institute of Physics, University of Silesia \\
Uniwersytecka 4, PL-40-007 Katowice, Poland \\

\vspace{2cm}
{\bf Abstract} \\
\end{center}
The production of heavy and light neutrinos in $e^+e^-$ future colliders
is considered. The cross section for the process $e^+e^- \rightarrow \nu N$
and then the  heavy neutrino decay $N \rightarrow W^{\pm} e^{\mp}$
is determined for experimentally possible values of mixing matrix elements.
The bound on the heavy neutrino-electron mixing is estimated in models
without right-handed currents. The role of neutrino CP
eigenvalues and the mass of the lightest Higgs particle are investigated.
The angular distribution of charged leptons in the total CM frame resulting
from the heavy neutrino decay and from the main $W^+W^-$ production background
process are briefly compared. 
\end{titlepage}
\vspace{0.5 cm}
\section{Introduction}
Our experimental knowledge about neutrinos is still relatively small. 
The results of 
terrestrial experiments agree with the prediction of the standard model (SM) 
where neutrinos are massless, left current
interacting particles. As a consequence we do not even  know if neutrinos have 
Dirac or Majorana character. There are however, astrophysical observations and 
cosmological estimations which, most probably, require massive neutrinos [1].
There is also the first terrestrial experiment in which there is some
indications
that a neutrino oscillates [2] and as a consequence at least one should be
massive. 
The existence of such small mass neutrinos is predicted by many
extensions of the SM. Usually the light neutrinos are accompanied by neutrinos 
with large mass in such a way that the so called see-saw mechanism [3]
occurs. 
The production of heavy neutrinos in the future linear colliders depends on 
their masses and couplings to known leptons and bosons.
The couplings of a neutrino below the $M_Z$ mass are strongly restricted by 
present LEP data [4] so we
will concentrate on neutrinos with masses above the $Z_0$ mass. 
If the explanation of small neutrino masses is given by the see-saw 
mechanism then the present experimental bounds for the light
(eV-keV-MeV region) and the heavy neutrinos $M_N>M_Z$ give very small mixing 
angles. With such mixing angles the heavy neutrinos decouple from low energy 
physics and the cross section for their
production in the future linear colliders is beyond our experimental interest. 
There are, however, models
where light-heavy neutrino mixings are not connected with the see-saw mechanism. 
The general idea can
be explained by an elementary example of the `light' $(\nu)$ and 
the `heavy' (N) neutrino. Let us assume that in the $\left( \nu, N \right)^T$ basis 
the neutrino mass matrix is
\begin{equation}
M=\left( \matrix{ a & b \cr b& c } \right),
\end{equation}
where for simplicity we assume that all elements a,b,c are real numbers. 
The masses and the mixing angle are given by
\begin{equation}
m_{1,2}=\frac{1}{2}\left(a+c\mp\sqrt{(a-c)^2+4b^2} \right),
\end{equation}
and
\begin{equation}
\sin{2\xi}=\frac{2b}{\sqrt{(a-c)^2+4b^2}}.
\end{equation}
There are two ways of predicting the light-heavy spectrum of neutrino masses. 
One is the see-saw mechanism where a=0,  $c>>b$ and then
\begin{equation}
\mid m_1 \mid \simeq \frac{b^2}{c},\;\;\; \mid m_2 \mid \simeq c>>m_1,
\end{equation}
and, unavoidably,
\begin{equation}
\xi \simeq \frac{b}{c} \simeq \sqrt{\frac{\mid m_1 \mid}{m_2}}<<1.
\end{equation}
The other one in which we assume that $a \neq 0$ and due to internal symmetry 
$ac=b^2$ gives
\begin{eqnarray}
m_1&=&0, \nonumber \\
m_2&=&a+c, 
\end{eqnarray}
and
\begin{equation}
\sin{\xi}=\frac{2\sqrt{ac}}{a+c}.
\end{equation}
If the symmetry, which at the tree level gives the relation $ac=b^2$, is broken 
we obtain
$$m_1 \neq 0 << m_2$$ in the higher order (see e.g. [5]). In this sort of models 
$\sin{2\xi}$ is not connected with the ratio $m_1/m_2$ and can be large 
$( \sin{2\xi} \simeq 1)$ for $a \simeq c$. Any model realizing this idea
in the natural way is an alternative to the see-saw mechanism and helps 
to explain the spectrum of neutrino
masses. Several kinds of such models were considered in literature [6]. 
In these scenarios the mixing angles are independent parameters not connected 
to the neutrino masses and are only bound by existing experimental data. 
In this paper we have found such boundary for mixing parameters  which
is model independent. We also assume that heavy neutrinos exist with such masses 
that they can be produced in future $e^+e^-$ colliders [7]. With these
assumptions we determined the cross section for the production of heavy and  
light neutrinos in future $e^+e^-$ colliders. We have also considered  the decay
of heavy neutrinos $N \rightarrow W^+l^-$ or $W^-l^+$ and the angular distribution 
of charged leptons in the total CM system. The decay channel is easily 
distinguished from the charged lepton production in various background
processes where $W^{\pm}$ pair production and decay are dominant. The effect
of the lightest, SM Higgs particle on the process $e^+e^- \rightarrow \nu N (W^+e^-\;
or\; W^-e^+)$ is also discussed. The process of production of heavy neutrinos
in $e^+e^-$ colliders has already been considered in literature [8]. However, to
our knowledge, the
analysis with all details mentioned above have not been performed.

In the next Chapter the bounds on mixing matrix elements using the full 
experimental information are given.
In Chapter 3  the angular distribution for final electron
(positron) in the  process
\arraycolsep0.5mm
\begin{equation}
e^+e^- \rightarrow  
\begin{array}[t]{ll}
\nu & N \qquad  \\
& \hookrightarrow  e^{\pm}W^{\mp}
\end{array}
\end{equation}
is calculated.
Conclusions are given in  Chapter 4. 

\section{Mixing matrix elements.}

The cross sections for production and decay of heavy neutrinos 
(Eq.~(8)) are given in the Appendix (Eqs.~A.1,A.2,A.3). The mixing
matrix elements $K_{Nl}$ and $\Omega_{N\nu}$ of the lepton sector analog of
Kobayashi-Maskawa matrices [9], decide about the magnitude
of the cross section. Precisely the helicity amplitudes are proportional to
\begin{eqnarray}
\left( K_{Ne} \right)^2 K_{\nu e} \;\;\;\;\;\;\;\;\;\;\;\;\;\;\;\;\;\;\;\;\;\;
\;\;\;\;\;\;\;\;\;\;\;\;\;\;\;\;\;\;\;
& \mbox{\rm in the t and u channels,} \nonumber \\
and \hspace{7 cm}\;\;\;\;\;\;\;\;\;\; && \nonumber \\
K_{Ne}\Omega_{N\nu}\;\;\;\; \mbox{\rm where } \;\;\; \Omega_{N\nu}=\sum\limits_
{l=e,\mu,\tau}K_{Nl} K_{\nu l}^{\ast}& \mbox{\rm in the s
channel.} 
\end{eqnarray}
From the present experimental data we are not able to determine all
elements of the K matrix. Fortunately, with good approximation  only
one mixing matrix element $K_{Ne}$ between electron and the lightest heavy
neutrino N will decide about the size of the cross section and  
it is possible to determine the bound on it from existing
experimental data.
Phases of $K_{Nl}$'s have no influence (there is no t-u interference) 
and this means that no  effects of CP violation are  seen in the process [10].

In the previous paper [11] we have analyzed the existing experimental data
which restrict the mixing matrix elements. Three
different combinations of light and heavy neutrino masses and their mixing with
leptons are  possible to be limited:
\\

(i) from the lack of lepton number violation processes (e.g. $\mu \rightarrow e 
\gamma,\mu \rightarrow 3e, \mu \rightarrow e$ conversion in nuclei [12]) and from
the number of light neutrino species $N_{\nu}$ it is possible to get
\begin{equation}
\sum_{N(heavy)} \mid K_{Ne} \mid^2 \leq \kappa^2,
\end{equation}
and the lack of a signal in neutrinoless double-$\beta$ decay 
$(\beta\beta)_{0\nu}$ gives two bounds \\

(ii) for the light neutrinos
\begin{equation}
\mid \sum_{\nu (light)}K_{\nu e}^2 m_{\nu} \mid < \kappa_{light}^2,
\end{equation}

(iii) for the heavy neutrinos
\begin{equation}
\mid \sum_{N(heavy)}K_{Ne}^2\frac{1}{m_N} \mid < \omega^2.
\end{equation}

The matrix K must be unitary and this means that \\

(iv) 
\begin{equation}
\sum_{\nu (light)} \mid K_{\nu e} \mid^2+\sum_{N (heavy)} 
\mid K_{N e} \mid^2=1.
\end{equation}

In paper [11] we have also used the constraints which follow from the
lack of Higgs triplets in considered gauge models. As a result, in the first 
order, the mass term for left-handed neutrinos does not appear. Here we 
will omit 
this assumption. In this way the limits which we get are
model independent. To find the inequalities (10)-(12) only one model assumption
is made, i.e. the lack of right-handed current hence our considerations are valid
for any model without right-handed charged currents. We know however, that
due to large mass of the right-handed gauge boson(s) $W_R^{\pm}$, the influence
of right-handed current on the production of one light and one heavy
neutrino is marginal [13].

Using restrictions (i)$\leftrightarrow$(iv) 
the upper bound on $K_{Nl}$ mixing depends on 1) the number of heavy
neutrinos ($n_R$) and 2) their CP parities ($\eta_{CP}$).

$\bullet$ $n_R=1$

For heavy neutrino with mass less than 1 TeV ($M<1$ TeV) we get from
relation (12)
\begin{equation}
\mid K_{Ne} \mid^2 < \omega^2 M
\end{equation}
and the total cross section is bounded by the small value of $\omega$
(see next Chapter).

$\bullet$ $n_R=2$

There are two heavy neutrinos with masses $M_1=M$ and $M_2=AM$ ($A \geq 1$).
The couplings depend on the CP parities of both neutrinos. If they are the
same e.g. $\eta_{CP}(N_1)=\eta_{CP}(N_2)=+i$ then mixing parameters can be 
treated as real
$K_{N_1e}=x_1$ and $K_{N_2e}=x_2$. The relations (10) and (11) give
\begin{eqnarray}
x_1^2+x_2^2 & \leq & \kappa^2, \nonumber \\
\mid x_1^2+\frac{x_2^2}{A} \mid & \leq & \omega^2 M,
\end{eqnarray}
and the situation is the same as in  case $n_R=1$ (Eq.~(14)), the coupling of the
$N_1$ neutrino is small $x_1^2 \leq \omega^2 M$.
If, however, heavy neutrinos have opposite CP parities $\eta_{CP}(N_1)=
-\eta_{CP}(N_2)=i$ then $K_{N_1e}=x_1,\;K_{N_2e}=ix_2$ and the relations
(10) and (12) give
\begin{eqnarray}
x_2^2 & \leq & \kappa^2-x_1^2, \nonumber \\
x_2^2 & \geq & A (x_1^2-\omega^2M), \nonumber \\
{\rm and} \;\;\;\;\;\;\;\;\;\; && \nonumber \\
x_2^2 & \leq & A (x_1^2+\omega^2M).
\end{eqnarray}
 The sketch of the region in $x_1^2 \leftrightarrow x_2^2$ plane of still
experimentally acceptable mixing parameters is shown in Fig.1. The maximum
value of $K_{N_1e}^2$ is equal
\begin{equation}
( K_{N_1e}^2 )_{max} = \frac{\kappa^2+AM\omega^2}{A+1}.
\end{equation}

\vspace{6 cm}
\begin{figure}[h]
\includegraphics{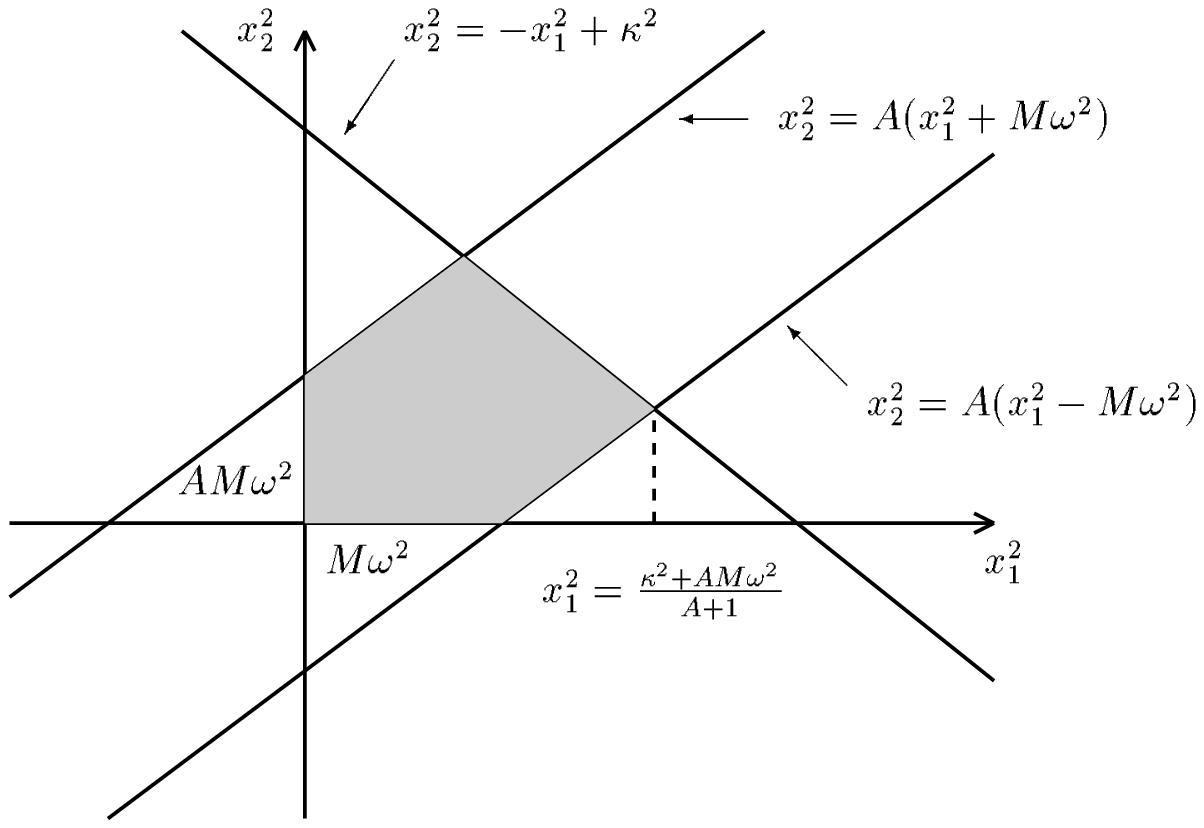}
\end{figure}
{\footnotesize Fig.1 Sketch of the region in $x_1^2 \leftrightarrow x_2^2$ plane of still
experimentally acceptable mixing parameters for two heavy neutrinos. Maximum 
value of $x_1^2$ is equal $(x_1^2)_{max}=\frac{\kappa^2+AM\omega^2}{A+1}$ and
approaches $\frac{\kappa^2+M\omega^2}{2}$ for $A \rightarrow 1$.}

\newpage

$\bullet$ $n_R=3$

If the CP parities of all neutrinos are the same 
$\eta_{CP}(N_1)=\eta_{CP}(N_2)=\eta_{CP}(N_3)=\pm i\;$
then all couplings are small and the same inequality (14) as in the $n_R=1,2$
cases restricts the $K_{N_1e}$ mixing.

A more interesting situation arises if we assume that not all $\eta_{CP}$'s
of neutrinos are the same. Let us assume that
$\eta_{CP}(N_1)=\eta_{CP}(N_2)=-\eta_{CP}(N_3)=+i$ then $K_{N_1e}=x_1,\;
K_{N_2e}=x_2$ and $K_{N_3e}=ix_3$. From relations (10) and (12) we obtain
three inequalities ($M_1=M,\;M_2=AM,\;M_3=BM$)
\begin{eqnarray}
x_3^2 & \leq & -x_1^2-x_2^2+\kappa^2, \nonumber \\
x_3^2 & \geq & Bx_1^2+\frac{B}{A}x_2^2-BM\omega^2, \nonumber \\
and \;\;\;\;\;\;\;\;\;\; && \nonumber \\
x_3^2 & \leq & Bx_1^2+\frac{B}{A}x_2^2+BM\omega^2.
\end{eqnarray}

The region in $(x_1^2,x_2^2,x_3^2)$ frame of still experimentally acceptable
parameters is shown in Fig.2. The maximum value of $K_{N_1e}^2$ is equal
\begin{equation}
(K_{N_1e}^2)_{max}=\frac{\kappa^2+BM\omega^2}{B+1}
\end{equation}
and can be as large as $(\kappa^2+M\omega^2)/2$ for $B \rightarrow 1$. \\

The other combination of $\eta_{CP}$'s leads to the bound on $K_{N_1e}^2$
which is the same as in the case $n_R=1$ or to this given by Eq.(19). So
finally we can state that regardless of the number of heavy neutrinos
the most optimistic bound on  $\mid K_{Ne} \mid^2$ is equal $\mid K_{Ne} 
\mid^2< \omega^2 M$ if there are no  correlations between elements of the K
matrix or $\mid K_{Ne} \mid^2 < (\kappa^2+\omega^2M)/2$ if there are 
correlations and some $\eta_{CP}$'s of heavy  neutrinos 
are opposite.

\newpage
\ \\

\vspace{8 cm}
\begin{figure}[ht]
\includegraphics{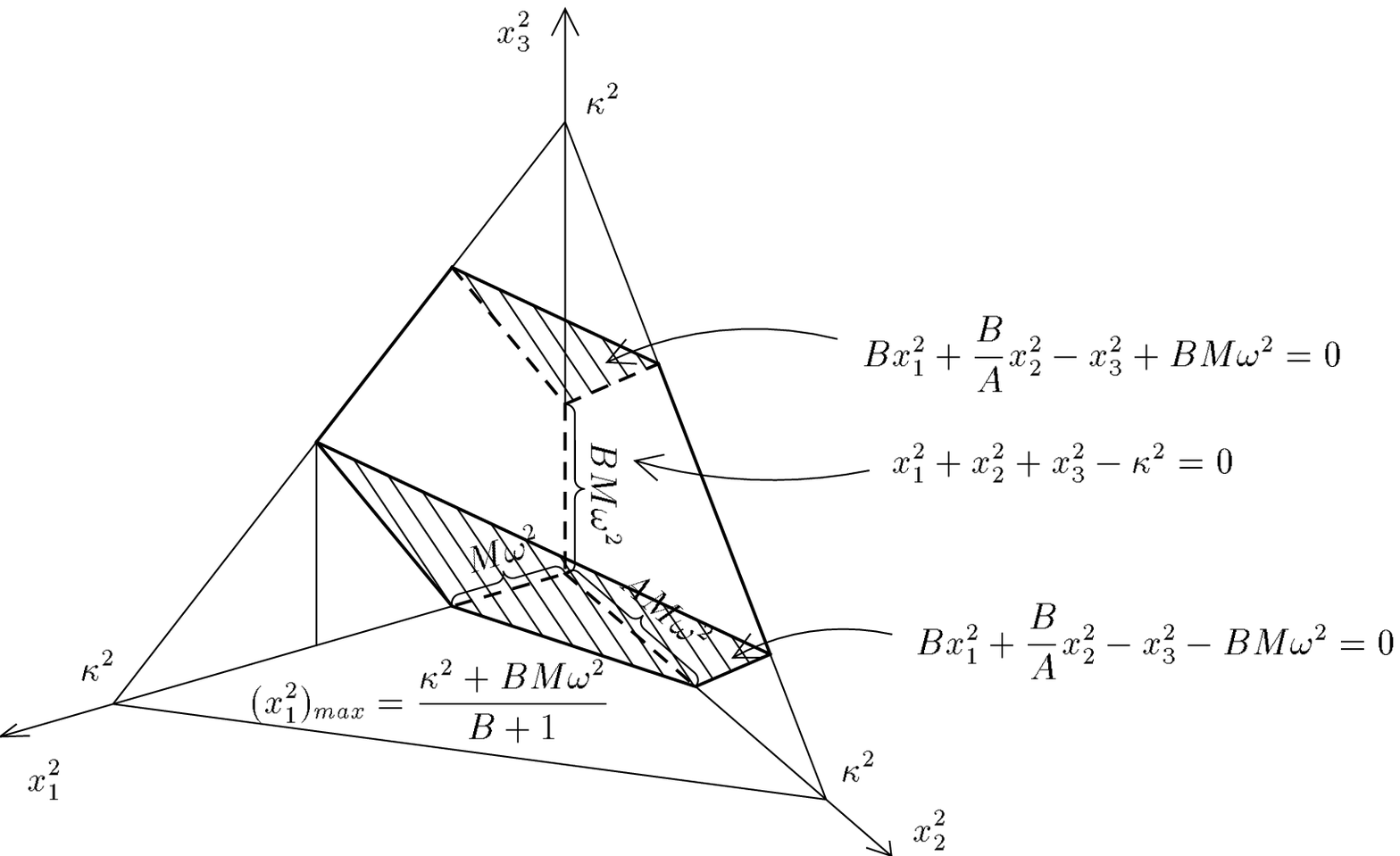}
\vspace{0.5 cm}
\end{figure}

{\footnotesize Fig.2 Sketch of the region in $(x_1^2,x_2^2,x_3^2)$ 
plane of still
experimentally acceptable mixing parameters for three heavy neutrinos
($n_R=3$). The region of acceptable parameters is bound by three reference
frame planes ($x_1^2,x_2^2$),($x_1^2,x_3^2$),($x_2^2,x_3^2$) and the
planes indicated in the Figure. The maximum value of $(x_1^2)$ is equal
$(x_1^2)_{max}=\frac{\kappa^2+BM\omega^2}{B+1}$ and approaches 
$\frac{\kappa^2+M\omega^2}{2}$ for $B \rightarrow 1$.}

\section{Numerical results}
\subsection{Production and decay of heavy neutrinos}
The light neutrinos will not be detected in the process $e^+e^- \rightarrow \nu N$
and  we can only measure  the sum
\begin{equation}
\sigma_{tot}=\sum\limits_{i=e,\mu,\tau} \sigma( e^+e^- \rightarrow \nu_i N),
\end{equation}
over all light neutrinos. For N we take the lightest heavy neutrino $N=N_1$. 
But from Eq.(A.1) (neglecting charged lepton masses)
\begin{eqnarray}
\sigma_{tot} & \propto & \mid K_{Ne} \mid^2 \left( \mid K_{\nu_e e}
\mid^2+\mid K_{\nu_{\mu}} \mid^2 + \mid K_{\nu_{\tau}} \mid^2 \right) 
\nonumber \\
&=& \mid K_{Ne} \mid^2 ( 1- \sum_N \mid K_{Ne} \mid^2 )^2
\simeq \mid K_{Ne} \mid^2.
\end{eqnarray}

To calculate the cross section $\frac{d\sigma}{d \cos{\Theta_e}}$ (Eq.(A.3))
we also need to
know the total decay width $\Gamma_N$ for heavy neutrino decay. From
Eqs. (A.2) we can calculate the partial decay width for 
$$N \rightarrow W^{\pm}l^{\mp}\;\;\;\; \mbox{\rm decay} $$
\begin{equation}
\Gamma(N \rightarrow W^{\pm}l^{\mp})=\frac{ \mid K_{Nl} \mid^2}{8 \sqrt{2} \pi}
\frac{G_F}{m_N^3}(m_N^2+2m_W^2)(m_N^2-m_W^2)^2,
\end{equation}
and $$N \rightarrow Z \nu_l \;\;\;\;\; \mbox{\rm decay} $$
\begin{equation}
\Gamma(N \rightarrow Z \nu_l)=\frac{ \mid \Omega_{N\nu_l} \mid^2}{8 \sqrt{2} \pi}
\frac{G_F}{m_N^3}(m_N^2+2m_Z^2)(m_N^2-m_Z^2)^2.
\end{equation}
Whether the decay channels of the N into the lightest Higgs particle H
and light neutrinos $\nu_l$, $N \rightarrow \nu_l H$ 
are opened depends on the relation between masses $m_N$ and $m_H$, if $m_N>m_H$
the channels are opened and (see e.g. [5])
\begin{equation}
\Gamma(N \rightarrow H \nu_l)=\frac{ \mid \Omega_{N \nu_l} \mid^2}{8 \sqrt{2} \pi}
\frac{G_F}{m_N}(m_N^2-m_H^2)^2.
\end{equation}
We will consider both situations where $m_N>m_H$ and  $m_N<m_H$ when the decay 
channel
is closed. However, since we are looking for a relatively light
$m_N$ ($\sim 100 \div 200$ GeV) 
the situation where $m_N<m_H$ (if $m_H \sim 300$ GeV) seems more plausible. 

The total decay width we calculate from
\begin{equation}
\Gamma_N=\sum_l \left( {2\Gamma(N \rightarrow l^+W^-)+ \Gamma(N \rightarrow \nu_l Z)
+\Gamma(N \rightarrow \nu_l H)\Theta(m_N-m_H)} \right)
\end{equation}
where
\begin{equation}
\sum_l \Gamma(N \rightarrow l^+W^-) \propto \sum\limits_{l=e,\mu,\tau}
\mid K_{Nl} \mid^2 \simeq \mid K_{Ne} \mid^2,
\end{equation}
\begin{equation}
\sum_l \Gamma(N \rightarrow \nu_l H),
\sum_l \Gamma(N \rightarrow \nu_l Z) \propto \sum_l
\mid \Omega_{N\nu_l} \mid^2 \simeq \sum_l \mid K_{Nl} \mid^2 \simeq
\mid K_{Ne} \mid^2.
\end{equation}
In the approximations made in Eqs. (21,26 and 27) we assume that in each column 
of K matrix ($l=e,\mu,\tau$)
$$(K_{\nu_el}, K_{\nu_{\mu}l}, K_{\nu_{\tau}l}, K_{N_1l}, K_{N_2l}, ...)^T$$
one element $K_{\nu_ll} \simeq 1$ (lepton universality) and only one coupling
between heavy neutrinos and lepton is visible $K_{Nl} \simeq x$. All other
couplings are very small and we neglect them.

The calculated decay width $\Gamma_N$ normalised to the factor $\mid K_{Ne} \mid^2$
for various masses $m_N$ is given in the 
Table 1. 

Now we have all the ingredients to calculate the electron angular distribution
in the process
$$ e^+e^- \rightarrow \nu N \rightarrow \nu e^-W^+.$$
In our approximation only one parameter $\mid K_{Ne} \mid^2$ decides about
the value of the cross section. For $n_R=1$, regardless  of the $\eta_{CP}$ 
of the heavy 
neutrino, and for $n_R>1$ with the assumption that $\eta_{CP}$'s of all neutrinos
are the same, $\mid K_{Ne} \mid^2$ is bounded by the lack of neutrinoless 
double $\beta$ decay (Eq.(14)). 
There are problems with estimating the role of heavy neutrinos in 
the $(\beta\beta)_{0\nu}$ process as the nuclear structure matrix elements
are calculated with limited accuracy [14]. The best limit is found 
from absence of neutrinoless double beta
decay in $^{76}Ge$ by Heidelberg-Moscow collaboration [15]
$$\omega^2 < 2 \cdot 10^{-5}\; \mbox{\rm TeV}^{-1}.$$ 
There are also other estimations of $\omega^2$. In  paper
[16] it was found that
$$\omega^2 < 2.8 \cdot 10^{-5}\; \mbox{\rm TeV}^{-1}.$$ 
In Table II we give the maximum values of $\sigma_{tot}
(e^+e^- \rightarrow \nu N)$ (Eq.20) for various heavy neutrino masses $m_N$
and different total energies $\sqrt{s}$. 
The value of $\omega^2$ decides about $\sigma_{tot}(max);
\; \sigma_{tot} \propto \omega^2$ and the values of the total cross section
for various $\omega^2$ can be easily obtained from the Table. 
As the maximum value of $\mid K_{Ne}
\mid^2$ is proportional to $m_N$ (see Eq.(14)) the cross section (Eq.(20))
increases with the heavy neutrino mass with the exception when
$m_N \rightarrow \sqrt{s}$ at the end of the phace space.

For $n_R>1$ and for different values of $\eta_{CP}$ of heavy neutrinos the bound
from $(\beta\beta)_{0\nu}$ (Eq.14) is not so crucial and $\mid K_{Ne} \mid^2$ can be
much larger (Eqs. (17) and (19)). In the both considered cases $n_R=2$ and $n_R=3$
the largest possible value is
\begin{equation}
\mid K_{Ne} \mid^2_{max} \rightarrow \frac{ \kappa^2+M[TeV]\omega^2}{2} 
\stackrel{M \leq 1\;TeV}{\longrightarrow} \frac{\kappa^2}{2} 
\end{equation}
for almost degenerate heavy neutrinos ($A \rightarrow 1$ for n=2, $A
>> B, B \rightarrow 1$ for n=3). In the case B=1 there are two Majorana
neutrinos with the same masses and opposite CP parities which form the Dirac
neutrino. In our studies, however, calculation of the cross section for Dirac
neutrino production is not performed.
Different values of $\kappa^2$ are found for
the model with singlet neutrinos:
$\kappa^2 < 0.015$ [17] and the more recent one $\kappa^2<0.0054$ [18]. 
If we use the recent LEP result for 
the number of light neutrino species $N_{\nu}=2.989 \pm 0.012$ [19] we obtain
$\kappa^2< 0.0055$, a value very close to the global fit given in [18].
In Table III the total cross section $\sigma( e^+e^- \rightarrow \nu N)$ for
various $m_N$ and $\sqrt{s}$ is presented. Results are given for 
$\kappa^2=0.0054$. Since $\sigma_{tot} \propto \mid K_{Ne} \mid^2$, values
of the $\sigma_{tot}$ for various $K_{Ne}$ can be easily obtained from this
Table.

In Fig. 3 we present the angular distribution for the final electron
$e^-e^+ \rightarrow \nu (N \rightarrow e^-W^+)$ for
various masses of heavy neutrino $m_N=100,150$ and 200 GeV calculated for
the 
maximum possible value of $\mid K_{Ne} \mid^2 \simeq \frac{\kappa^2}{2}$. For 
$\kappa^2$ we take the value $\kappa^2=0.0054$.
Results are given for the Next Linear Collider
with CM energy $\sqrt{s}=500$ GeV. This distribution has forward-backward
symmetry. To show the influence of Higgs particle we present  results for 
$m_H=100$ GeV on the
left side of the Figure  $(-1 \leq \cos{\Theta_e} \leq 0)$
and on the right side $(0 \leq \cos{\Theta_e} \leq 1)$
the Higgs decay channels are excluded. For higher Higgs mass the total width $\Gamma_N$ is
smaller and, due to the greater value of the branching ratio for the $N 
\rightarrow lW$
decay, the cross section $\frac{d\sigma}{d \cos{\Theta_e}}$ is larger. 
Numerically, Higgs has no influence on the cross section for $m_N=100$ GeV
(for $m_H \geq 100$ GeV the $N \rightarrow \nu H$ decay channel is closed)
and the influence of the Higgs particle ($m_H=100$ GeV) is
approximately equal 10 \%, 15\% for $m_N=150,200$ GeV, respectively.
For higher energies the final electron distribution is more peaked
in the forward-backward direction $( \cos{\Theta_e}=\pm1)$. This is the result
of $W^{\pm}$ exchange in t and u channels and small
contribution of the s channel $Z^0$ exchange. For $\sqrt{s}=0.5$ TeV the
$Z^0$ exchange mechanism gives only 2\% contribution to the total cross section
[12] and is smaller for higher energies. As an
example we compared final electron distribution produced by the decay of a heavy
neutrino with mass $M_N=100$ GeV for $\sqrt{s}=500$ and 1000 GeV (Fig. 4).

\newpage
\ \\
\vspace{8 cm}
\begin{figure}[h]
\includegraphics{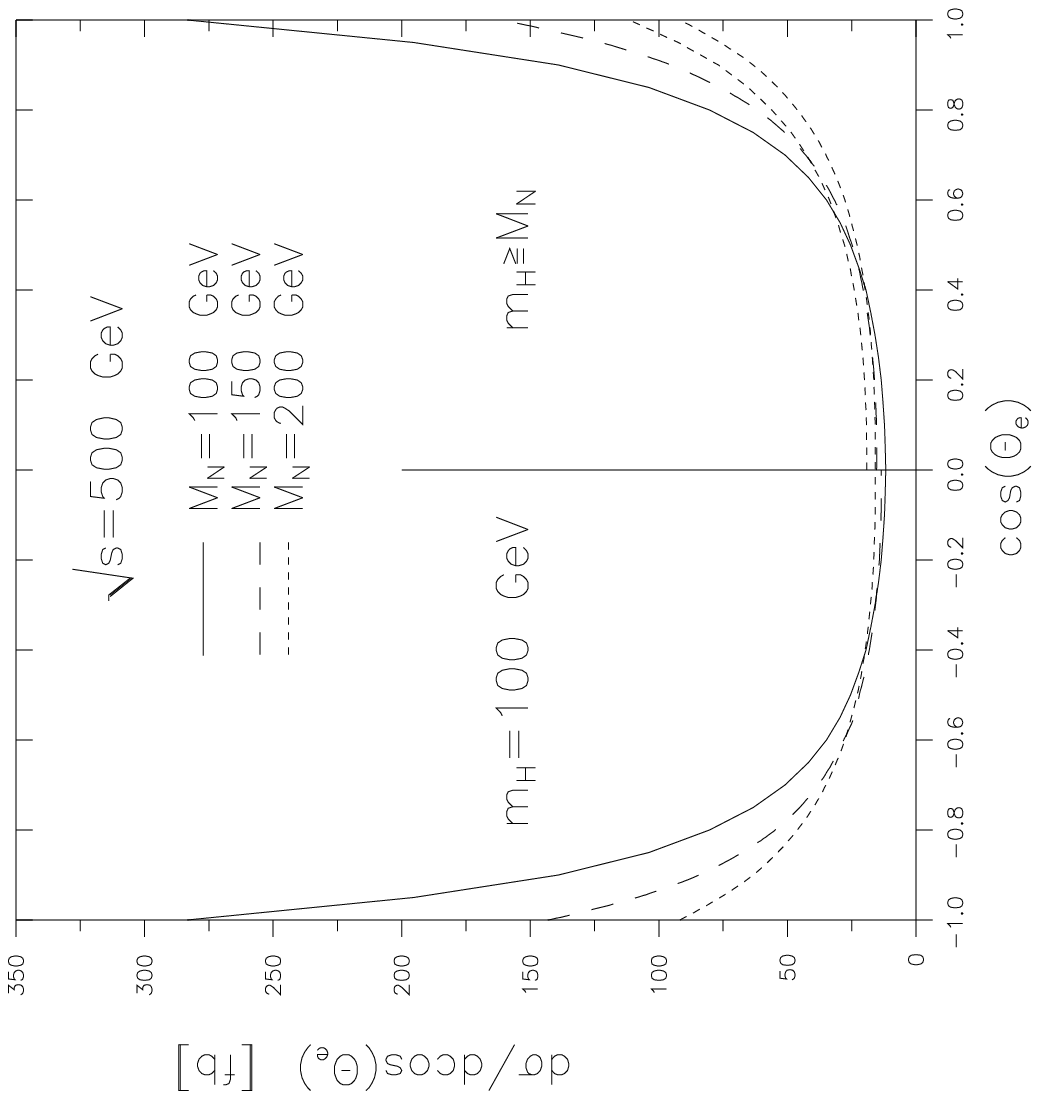}
\vspace{0.5 cm}
\end{figure}
\newline 

{\footnotesize Fig.3 Distribution of the final electron from a heavy neutrino decay for 
$\sqrt{s}=500$ GeV collider with $M_N=100$ GeV (solid line), $M_N=150$ GeV
(long-dashed line) and $M_N=200$ GeV (short-dashed line). Left
half of the Figure gives results for $m_H=100$ GeV. On the right-hand side
the Higgs decay channels are excluded.}

\newpage 
\ \\

\vspace{8 cm}
\begin{figure}[h]
\includegraphics{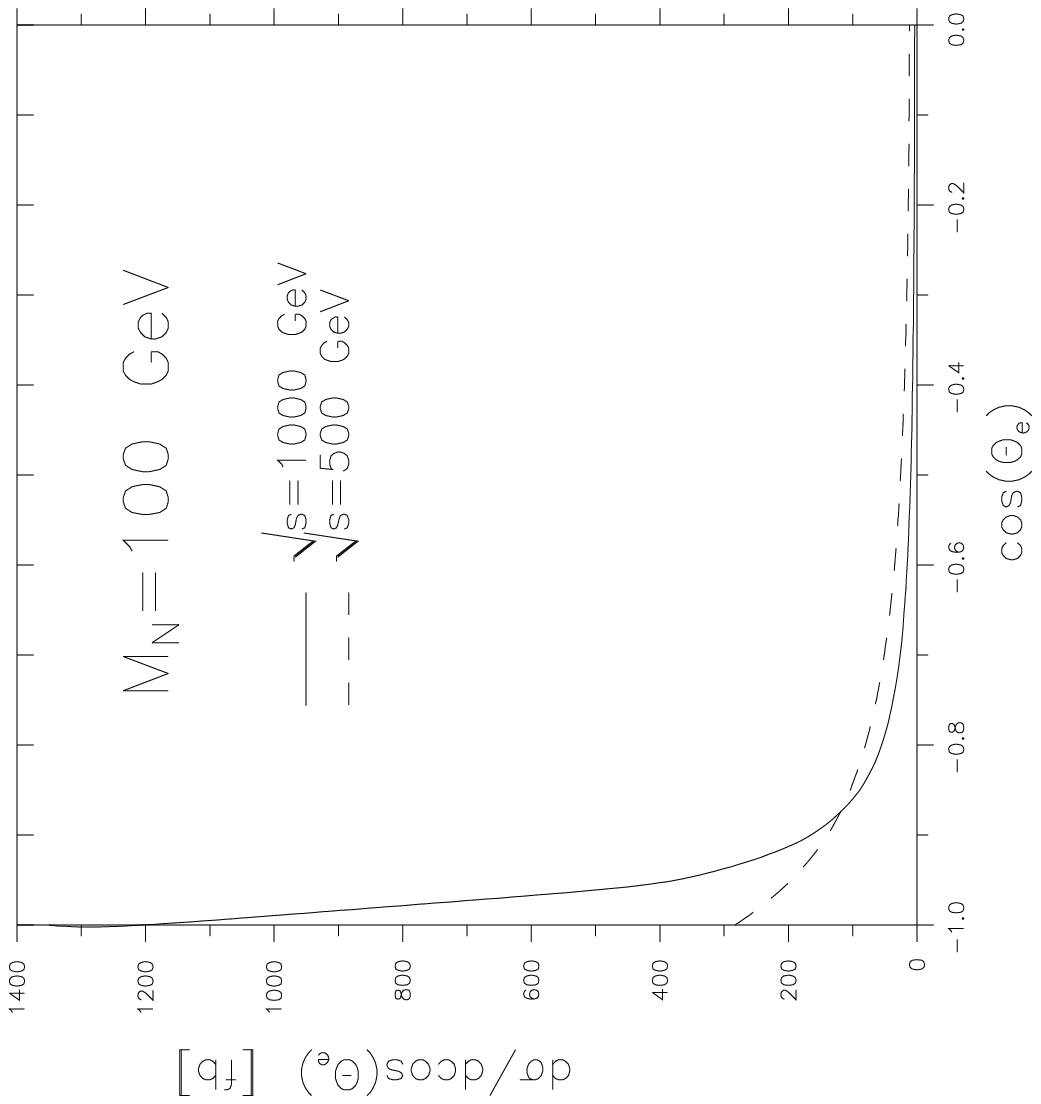}
\vspace{0.5 cm}
\end{figure}

{\footnotesize Fig.4 Backward distribution of the final electron 
coming from a heavy neutrino
decay ($M_N=100$ GeV) for two different energies: $\sqrt{s}=500$ GeV
(dashed line) and $\sqrt{s}=1000$ GeV (solid line). Forward distribution is 
the same.}
\ \\

Finally in Fig.5 we present the angular distribution $\frac{d\sigma}
{d\cos{\Theta_e}}$ for various masses of heavy neutrino $M_N=100,300$ and
500 GeV (for $m_H=100$ GeV). The cross section becomes higher and more
peaked in the forward-backward direction for smaller mass of heavy neutrinos.
The effect of growing $\frac{d\sigma}{d \cos{\Theta_e}}$ is the result
of increasing $BR(N \rightarrow lW)$ and increasing of $\sigma_{tot}
(e^+e^- \rightarrow \nu N)$ (Table III) for smaller $m_N$. The effect
of slope reducing with $m_N$ mass in the forward-backward direction is also
kinematically understandable.
\newpage
\ \\
\vspace{8 cm}
\begin{figure}[h]
\includegraphics{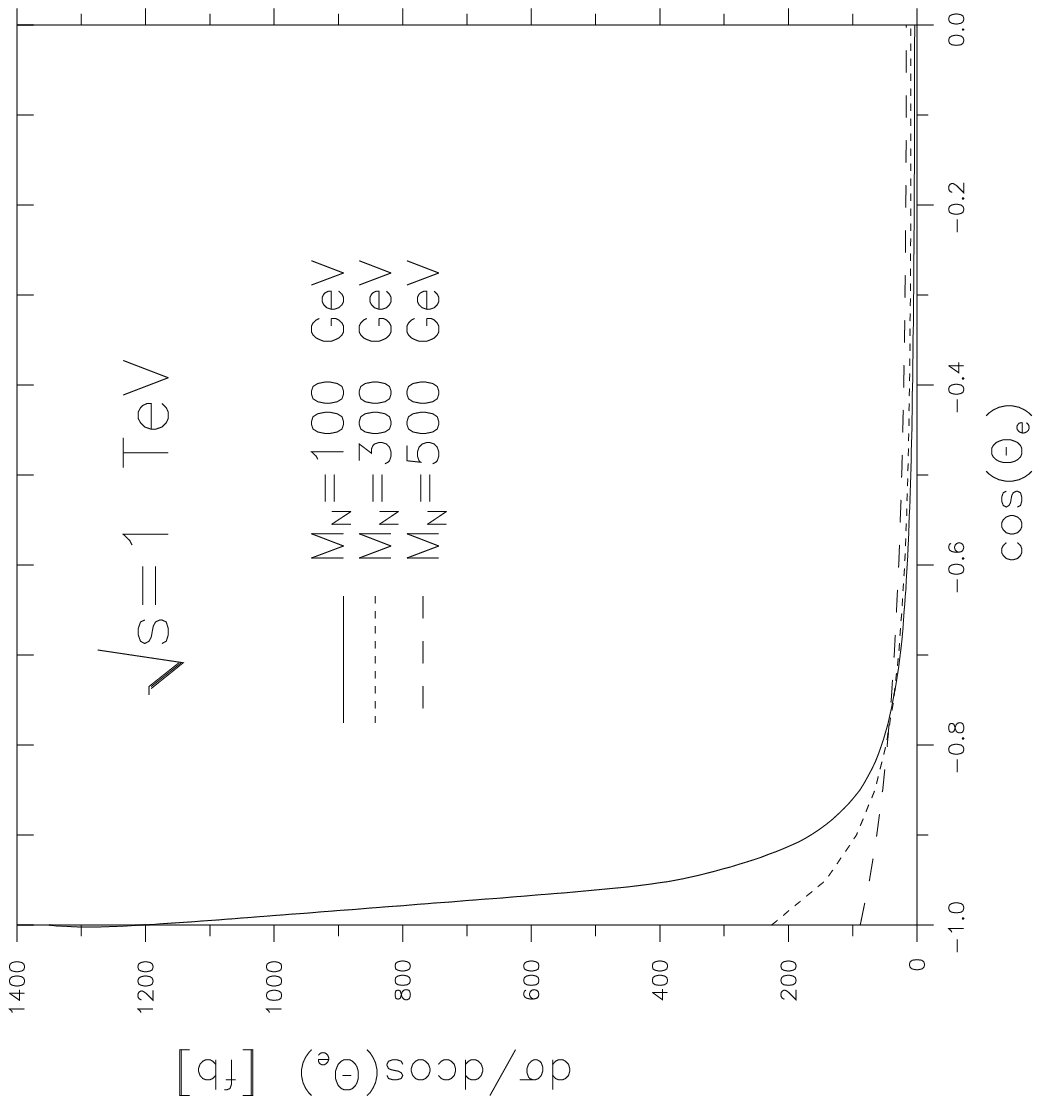}
\vspace{0.5 cm}
\end{figure}

{\footnotesize Fig.5 Backward distribution of the final electron coming from
a heavy neutrino
decay with mass $M_N=100$ GeV (solid line), $M_N=300$ GeV (short-dashed line)
and $M_N=500$ GeV (long-dashed line) for $\sqrt{s}=1$ TeV. Forward
distribution is the same.}
\ \\

The main background process is the production of $W^+W^-$ pair and then the
$W^{\pm} \rightarrow e^{\pm} \nu$ decay.
The distribution of charged lepton coming from the heavy neutrino decay (N)
already mentioned in this paper and from $W$'s decays by  
$e^+e^- \rightarrow W^+W^-$ process differs very much in forward-backward
direction.  
For high energy ($\sqrt{s}>0.5$ TeV) angular distribution of electrons
coming from the $W^-$ decay is peaked in the forward direction and has
reducing slope in background direction. On the 
contrary,
the $e^-$ coming from N decay will travel equally well both in the forward 
and the backward direction with increasing slope of angular distribution
for $| \cos{\Theta_e} | \rightarrow 1$ (Figs.3-5).

\section{Conclusions}

We have found the cross section for heavy and light neutrino production in
future electron-positron colliders with energy $\sqrt{s} \geq 0.5$ TeV. The
bounds on mixing matrix element $K_{Ne}$ between heavy neutrino and electron
are found from existing experimental data in models without right-handed
currents. The maximum possible value of the $K_{Ne}$ is very small if there
is only one heavy neutrino ($n_R=1$) or, in the case of a larger number of
heavy neutrinos ($n_R>1$), if their CP eigenvalues are the same. This small
bound results from the lack of a signal in neutrinoless double-$\beta$ decay.
In this case the cross section for production of light and heavy neutrinos 
($e^+e^- \rightarrow \nu N$) is very small from 0.16 fb for $\sqrt{s}=0.5$
TeV and $m_N=100$ GeV up to 1.6 fb for $\sqrt{s}=2$ TeV and $m_N=1$ TeV.

The lack of any signal from neutrinoless double-$\beta$ decay does not give
such a restrictive bound if the CP eigenvalues of two or more  
heavy neutrinos are
not the same. Now the $e^+e^- \rightarrow \nu N$ cross section can be larger
and equals $\sigma=240(287)$ fb for $\sqrt{s}=0.5(2)$ TeV and $m_N=100$ GeV.
We have also found angular distribution of the final charged lepton in the
total CM frame resulting from the heavy neutrino decay. The angular
distribution has forward-backward symmetry, contrary to background
process e.g. $e^+e^- \rightarrow W^+W^-(\rightarrow e^-\nu_e)$. This
property could point to the existence of a heavy neutrino. The charged
lepton angular distribution depends on CM energy, mass of the heavy
neutrino and mass of the lightest Higgs boson.

\section*{Appendix}
\setcounter{equation}{0}
\renewcommand{\theequation}{A.\arabic{equation}}
We would like to present the cross section for production ($e^+e^- \rightarrow
\nu N$) and decay of Majorana neutrino $N \rightarrow l^{\pm}W^{\mp},\nu'Z$ 
processes which are very useful in practical application. We consider
the $Ne^-W^+$ interaction without the right-handed coupling and neglected 
the electron mass ($m_e=0$). \\

\underline{Production process $e^+e^- \rightarrow \nu N$}

The production process $e^-(\sigma)+e^+(\bar{\sigma}) \rightarrow \nu(\lambda)+
N(\bar{\lambda})$ is described by 8 helicity amplitudes ($\Delta\sigma=
\sigma-\bar{\sigma},\Delta\lambda=\bar{\lambda}-\lambda$)

\begin{equation}
M(\Delta\sigma;\lambda,\bar{\lambda})=\left( \sqrt{2} \right)^{1+ \mid 
\Delta\lambda \mid } \left\{ \frac{A_t}{t-M_W^2}-\frac{A_u}{u-M_W^2}+
\frac{A_s}{s-M_Z^2+iM_Z\Gamma_Z} \right\} D_{\Delta\sigma,\Delta\lambda}^
{1\;\;\ast} \left( \phi, \Theta, 0 \right),
\end{equation}
where $A_{t,u,s}$ are functions of fermion helicities
\begin{eqnarray*}
A_t(\Delta\sigma,\lambda, \bar{\lambda})&=&K_{Ne}^{\ast}K_{\nu e}\sqrt{1+2
\bar{\lambda}\beta}\delta_{\lambda=-1/2}\delta_{\Delta\sigma=-1}, \\
A_u(\Delta\sigma,\lambda, \bar{\lambda})&=&K_{Ne}K_{\nu e}^{\ast}\sqrt{1-2
\bar{\lambda}\beta}\delta_{\lambda=+1/2}\delta_{\Delta\sigma=-1},  \\
A_s(\Delta\sigma,\lambda, \bar{\lambda})&=&\left[ \frac{1}{2}(-1+2\tan^2{\Theta_W})
\delta_{\Delta\sigma=-1}+
\tan^2{\Theta_W}\delta_{\Delta\sigma=+1} \right] \\
&\times& \left[ \Omega_{N\nu}\sqrt{1+2\bar{\lambda}\beta}\delta_{\lambda=-1/2}-
\Omega_{N\nu}^{\ast}\sqrt{1-2\bar{\lambda}\beta}\delta_{\lambda=+1/2} \right],
\end{eqnarray*}
and
$$\beta=\frac{s-s_N}{s+s_N}.$$
s,t,u are ordinary Mandelstam variables; $\Theta$ and $\phi$ are CM azimuthal 
and polar angles of the heavy Majorana neutrino N with respect to the initial 
electron,
$\sqrt{s_N}$ is the invariant mass of the heavy neutrino, $\Theta_W$
is the Weinberg angle. \\

\underline{Decay process $N \rightarrow l^{\pm}W^{\mp},\nu Z$}

In the helicity rest frame of N ($\Theta_e^{\ast}$ and $\phi_e^{\ast}$ are
$l^{\pm}$'s or $\nu$'s
azimuthal and polar angles respectively) the decay process $N(\bar{\lambda}) \rightarrow
V(\lambda_V)+f(\lambda_f)$ is described by 4 helicity amplitudes (the final fermion
mass is neglected, $M_V$ is the gauge boson mass)
\begin{eqnarray}
T(\bar{\lambda};\lambda_V;\lambda_f)=\sqrt{s_N-M_V^2}F_{\lambda_V
\lambda_f}{D_{\bar{\lambda},\lambda_f-\lambda_V}^{1/2\;\; \ast}} (\phi_f^{\ast},
\Theta_f^{\ast},0)
\end{eqnarray}
where
\begin{eqnarray*}
F_{++}&=&\sqrt{2}X,\;\;F_{--}=\sqrt{2}Y,\;\;F_{0+}=\frac{\sqrt{s_N}}{M_V}X, \\
F_{0-}&=&\frac{\sqrt{s_N}}{M_V}Y,\;\;F_{+-}=F_{-+}=0
\end{eqnarray*}
and
$$
\left\{
\begin{array}{lll}
X=-\frac{e}{\sqrt{2}\sin{\Theta_W}}K_{Ne},\;& Y=0\;\; & \mbox{\rm for}\;
N \rightarrow W^-l^+, \cr 
X=0,\;\; & Y=\frac{e}{\sqrt{2}\sin{\Theta_W}}K_{Ne}^{\ast},\;\;\; & 
\mbox{\rm for}\; N \rightarrow W^+l^-, \cr 
X=-\frac{g}{2\sin{\Theta_W}\cos{\Theta_W}}\Omega_{N\nu},\; &
Y=\frac{g}{2\sin{\Theta_W}\cos{\Theta_W}}\Omega_{N\nu}^{\ast},\; 
& \mbox{\rm for} N \rightarrow \nu Z.
\end{array}
\right.
$$ 
\\

\underline{Full cross section}

The angular distribution of the final lepton in the $e^+e^-$ CM frame in the 
process $e^+e^- \rightarrow \nu N(\rightarrow e^{\pm}W^{\mp})$ is given by
$\large($ ($\Theta_e,\phi_e)$ are the CM azimuthal and polar angles of final
$e^{\pm}$ with respect to the initial electron $(e^-) \large)$

\begin{eqnarray}
\frac{d\sigma}{d\cos{\Theta_e}}&=&
\frac{G_F^2M_W^2}{ 2^{14}s^2\pi^5} \int_0^{2\pi}d\phi \int_{-1}^1
d\cos{\Theta}\int_0^{2\pi}d\phi_e \int_{M_W^2}^sds_N \nonumber \\
&&{\bf J} \frac{(s_N-M_W^2)(s-s_N)}{s_N \left[ (s_N-m_N^2)^2+M_N^2
\Gamma_N^2 \right] } \nonumber \\
&&\sum\limits_{\Delta\sigma;\lambda,\lambda_V,\lambda_f} \mid 
\sum_{\bar{\lambda}}M\left( \Delta\sigma;\lambda,\bar{\lambda} \right)
T\left( \bar{\lambda};\lambda_V,\lambda_f \right) \mid^2 \nonumber
\end{eqnarray}
\begin{equation}
\end{equation}

where ${\bf J}$ is the Jacobian of the transformation between the $e^{\pm}$ 
angles in the
CM frame of the decaying neutrino and the CM frame of initial colliding leptons
\begin{eqnarray}
{\bf J}&=&\frac{1-\beta^2}{(1-\beta z)^2w} \left\{ \sin^2{\Theta}\sin^2{\left(
{\phi}_e+\phi \right)} \right. \nonumber \\
&+&\left. \left( \cos{\Theta}\sin{{\Theta}_e}-
\sin{\Theta}\cos{{\Theta}_e}\cos{\left( {\phi}_e+\phi \right) }
\right)^2 \right\}
\end{eqnarray}
where
\begin{eqnarray}
w &=&\sin^2{\Theta_e}\sin^2{\left(
\phi_e+\phi \right)} \nonumber \\
&+&\left(\cos{\Theta}\sin{\Theta_e} \cos{ \left( \phi_e+\phi \right)}-
\sin{\Theta}\cos{\Theta_e} \right)^2, \\
\mbox{\rm and} && \nonumber \\
z&=&\sin{\Theta}\sin{\Theta_e} \cos{ \left( \phi_e+\phi \right) }+
\cos{\Theta}\cos{\Theta_e} .
\end{eqnarray}

The amplitude $T(\bar{\lambda},\lambda_V,\lambda_f)$ in Eq.(A.2) is introduced
in the CM frame of the decaying neutrino. We need to determine the exact dependence
between $\Theta_e,\phi_e$ and $\Theta_e^{\ast},\phi_e^{\ast}$ variables.
They are given by the relations

\begin{equation}
\cos{\Theta_e^{\ast}}=\frac{-\beta+z}{1-\beta z},
\end{equation}

\begin{equation}
\tan{\phi_e^{\ast}}=\frac{\sin{\Theta_e}\sin{\left(\phi_e-\phi
\right)}}{  \cos{\Theta}\sin{\Theta_e} \cos{( \phi_e-\phi )}-
\sin{\Theta}\cos{\Theta_e}} 
\end{equation}
and
\begin{equation}
sign(\sin{\phi_e^{\ast}})=sign(\sin{(\phi_e+\phi)}),
\end{equation}

($\tan{\phi_e^{\ast}}$ and $sign(\sin{\phi_e^{\ast}})$ describe the 
$\phi_e^{\ast}$ univocally in the region $0<\phi_e^{\ast}<2 \pi$).

\section*{Acknowledgements}
This work was partly supported by the Polish Committee for Scientific Research
under Grant No.~PB 659/P03/95/08 and by the Curie Sk\l odowska grant MEN/NSF
93-145.

\section*{References}
\newcounter{bban}
\begin{list}
{$[{\ \arabic {bban}\ }]$}{\usecounter{bban}\setlength{\rightmargin}{
\leftmargin}}
\item G.Gelmini, E.Roulet, Rep. Prog. Phys. {\bf 58} (1995) 1207.
\item The LSND Collaboration, C.~Athanassopoulos et al., \newline 
Phys.Rev.Lett. {\bf75}(1995)2650; ibid. {\bf 77}(1996)3082.
\item T.~Yanagida, Prog.~Theor.~Phys. {\bf B135} (1978) 66; M.~Gell-Mann,
P.~Ramond and R.~Slansky, in `Supergravity', eds. P.~Nieuwenhuizen and
D.~Freedman (North-Holland, Amsterdam, 1979) p.315.
\item L3 Collaboration, O.~Adriani et al., Phys. Lett. {\bf B295} (1992) 371
and {\bf B316} (1993) 427.
\item R.N. Mohapatra, P.B. Pal, "Massive neutrinos in physics and astrophysics",
World Scientific, 1991.
\item D. Wyler and L. Wolfenstein, Nucl. Phys. {\bf B218} (1983) 205;
R.N.~Mohapatra and J.W.F.~Valle, Phys. Rev. {\bf D34} (1986) 1642;
E.~Witten, Nucl. Phys. {\bf B268} (1986) 79;
J.~Bernabeu et al., Phys. Lett. {\bf B187} (1987) 303;
J.L.~Hewett and T.G.~Rizzo, Phys. Rep. {\bf 183} (1989) 193;
P.~Langacker and D.~London, Phys. Rev. {\bf D38} (1988) 907; E.~Nardi,
Phys. Rev. {\bf D48} (1993) 3277; D.~Tommasini, G.~Barenboim, J.~Bernabeu and
C.~Jarlskog, Nucl. Phys. {\bf B444} (1995) 451.
\item R.~Palmer, "Future accelarators", plenary talk at ICHEP, Warsaw 1996.
\item {\bf (a):} For LEPI energy and below the process $e^+e^- 
\rightarrow \nu N$
was studied before, see e.g. A.~Ali, Phys.Rev.{\bf D10} (1974) 2801;
M.~Gourdin, X.Y.~Pham, Nucl. Phys. {\bf B164} (1980) 387; J.L.~Rosner,
Nucl. Phys. {\bf B248} (1984) 503; M.~Ditmar, A.~Santamaria, 
M.C.~Gonzales-Garcia and J.W.F.~Valle, Nucl. Phys. {\bf B332} (1990)1;
M.C.~Gonzales-Garcia, A.~Santamaria and J.W.F.~Valle, Nucl. Phys. 
{\bf B342} (1990) 108; J.~Kugo and S.Y.~Tsai Prog. Theor. Phys. {\bf 86}
(1991) 183; J.W.F.~Valle Nucl.Phys.Proc.Suppl. {\bf 48}(1996)137 and 
hep-ph/9603307; A.~Hoefer and
L.M.~Sehgal, Phys.Rev. {\bf D54}(1996)1944;\newline
{\bf (b):} above LEPI energy, see e.g.
F.~del~Aguila, E.~Laermann and P.~Zerwas, Nucl. Phys. {\bf B297} (1988)1;
E.~Ma and J.~Pantaleone, Phys. Rev. {\bf D40} (1989) 2172; W.~Buchm{\"u}ller
and C.~Greub, Nucl. Phys. {\bf B363} (1991) 349 and {\bf B381} (1992) 109; 
J.~Maalampi, K.~Mursula and R.~Vuopionper{\"a}, Nucl. Phys. {\bf B372} (1992)23; 
M.C.~Gonzales-Garcia,
O.J.P.~Eboli, F.~Halzen and S.F.~Noaves Phys. Lett. {\bf B280} (1992) 313;
R.~Vuopionper{\"a} Z. Phys. {\bf C65} (1995) 311.
\item see e.g. J.~Gluza and M.~Zra\l ek, Phys. Lett. {\bf B362} (1995) 148.
\item J.~Gluza and M.~Zra\l ek, Phys. Rev. {\bf D51} (1995) 4707.
\item J.~Gluza and M.~Zra\l ek, Phys. Lett. {\bf B372} (1996) 259.
\item B.W.~Lee, R.~Shrock, Phys.Rev. {\bf D16}(1977)1444; B.W.~Lee,
S.~Pakvasa, R.~Shrock, H.~Sugawara, Phys.Rev.Lett. {\bf 38}(1977)937;
W.~Marciano, A.I.~Sanda, Phys.Lett. {\bf B37}(1977)303; T.P.~Cheng, L.F.~Li,
Phys.Rev. {\bf D44}(1991)1502.
\item J.Gluza and M.Zra\l ek, Phys.Rev. {\bf D48} (1993) 5093.
\item C.A.~Heusch and P.~Minkowski, hep-ph/9611353.
\item A.~Balysh et.~al., Phys. Lett. {\bf B356} (1995) 450.
\item T.~Bernatowicz et.~al., Phys. Rev. Lett. {\bf 69} (1992) 2341.
\item E.Nardi, E.Roulet and D.Tommasini, Nucl. Phys. {\bf B386} (1992) 239;
A.~Ilakovac and A.~Pilaftsis, Nucl. Phys. {\bf B437} (1995) 491.
\item A.Djoudi, J.Ng and T.G.Rizzo, hep-ph/9504210.
\item A.~Blondel, "Status of the electroweak interactions", plenary talk at
ICHEP, Warsaw 1996. 
\end{list}
\newpage

\begin{table}[h]
\begin{center}
\vspace{ 1cm}
\begin{tabular}{|c| c| c|} 
\cline{1-3}
\cline{1-3}
& \multicolumn{2}{|c|}{ }  \\ 
$M_N$ [GeV] & \multicolumn{2}{|c|}{ $\Gamma_N^{total}/\mid K_{Ne} \mid^2$   [GeV] }  \\ 
& \multicolumn{2}{|c|}{ }  \\ 
&& \\
& $m_H=100$ GeV & $m_H \geq m_N$ GeV  \\ 
&& \\
\cline{1-3}
\cline{1-3}
&& \\
100 & 0.22 & 0.22 \\ 
&& \\
150 & 2.9 & 2.6 \\
&& \\
200 & 8.7 & 7.2 \\
&& \\
300 & 33.1 & 26.1 \\
&& \\
500 & 160.2 & 143 \\
&& \\
700 & 445.5 & 337.5 \\
&& \\
1000 & 1306 & 984 \\
\cline{1-3}
\cline{1-3}
\end{tabular}
\end{center}
\end{table}
{\footnotesize {\bf Table 1.} The total width for a heavy neutrino decay divided by mixing matrix
element $\mid K_{Ne} \mid^2$ with the decay channels
$\Gamma \left( N \rightarrow \nu_l H \right) $ (second column) and without 
these channels
(third column) for various heavy neutrino masses $m_N$.} 

\newpage
\begin{table}
\begin{center}
\vspace{ 0.5cm}
\begin{tabular}{|c| c| c| c|} 
\cline{1-4}
\cline{1-4}
& \multicolumn{3}{|c|}{ }  \\ 
$M_N$ [GeV] & \multicolumn{3}{|c|}{ $\sigma^{total}_{max}$ [fb], $n_R=1$ }  \\ 
& \multicolumn{3}{|c|}{ }  \\ 
&&& \\
& $\sqrt{s}=0.5$ TeV & $\sqrt{s}=1$ TeV & $\sqrt{s}=2$ TeV \\ 
&&& \\
\cline{1-4}
\cline{1-4}
&&& \\
100 & 0.18 & 0.2 & 0.2 \\ 
&&& \\
150 & 0.25 & 0.3 & 0.3 \\
&&& \\
200 & 0.31 & 0.4 & 0.4 \\
&&& \\
300 & 0.34 & 0.6 & 0.6 \\
&&& \\
500 & - & 0.8 & 1.0 \\
&&& \\
700 & - & 0.7 & 1.3 \\
&&& \\
1000 & - & - & 1.6 \\
\cline{1-4}
\cline{1-4}
\end{tabular}
\end{center}
\end{table}
{\footnotesize {\bf Table 2.} Total cross section $\sigma_{tot} 
\left( e^+e^- \rightarrow \nu N \right) $ in $n_R=1$ case (see Eq.(14) with 
$\omega^2=2 \cdot 10^{-5}\;TeV^{-1}$)
for various heavy neutrino masses and three different total energies $\sqrt{s}=
0.5,1,2$ TeV. If $\omega^2 \simeq 80\cdot 10^{-5}\;TeV^{-1}$ [13] all
numbers in the Table should be multiplied by 40.}
\newpage
\begin{table}
\begin{center}
\vspace{ 0.5cm}
\begin{tabular}{|c| c| c| c|} 
\cline{1-4}
\cline{1-4}
& \multicolumn{3}{|c|}{ }  \\ 
$M_N$ [GeV] & \multicolumn{3}{|c|}{ $\sigma^{tot}_{max}$ [fb] $n_R>1$} \\
& \multicolumn{3}{|c|}{ }  \\ 
&&& \\
& $\sqrt{s}=0.5$ TeV & $\sqrt{s}=1$ TeV & $\sqrt{s}=2$ TeV \\ 
&&& \\
\cline{1-4}
\cline{1-4}
&&& \\
100 & 240 & 275 & 287 \\ 
&&& \\
150 & 227 & 271 & 286 \\
&&& \\
200 & 209 & 267 & 285 \\
&&& \\
300 & 155 & 252 & 281 \\
&&& \\
500 & - & 207 & 270 \\
&&& \\
700 & - & 138 & 252 \\
&&& \\
1000 & - & - & 216 \\
\cline{1-4}
\cline{1-4}
\end{tabular}
\end{center}
\end{table}
{\footnotesize {\bf Table 3} 
Total cross section $\sigma_{tot} \left( e^+e^- \rightarrow \nu N 
\right)$
for various heavy neutrino masses and total energies $\sqrt{s}$ calculated with largest
possible value of $\mid K_{Ne} \mid^2$ ($n_R>1$ case, see Eq.(28)). 
Result is given for $\kappa^2=0.0054$.} 
\end{document}